\documentclass[twocolumn,showpacs,preprintnumbers,amsmath,amssymb,pre]{revtex4}
\usepackage{graphicx}
\usepackage{dcolumn}
\usepackage{bm}

\usepackage{color}
\newcommand{\rev}[1]{#1}
\newcommand{\revb}[1]{#1}

\begin{document}
 
\title{Density profiles of loose and collapsed \rev{cohesive}  granular
  structures \revb{generated by ballistic deposition}}

\author{Dirk Kadau}
\author{Hans J. Herrmann}
 \affiliation{IfB, HIF E12, ETH H\"onggerberg,
8093 Z\"urich, Switzerland}

\date{\today}

\begin{abstract}
Loose granular structures stabilized against gravity by an effective cohesive
force are investigated on a microscopic basis using 
contact dynamics. We study the influence of the granular Bond
number on the density profiles and the generation process of packings,
generated by ballistic deposition 
 under gravity. The internal compaction occurs
discontinuously in small avalanches and we study their size distribution. We
also 
develop a model explaining the final density profiles based on insight about the collapse of a packing under changes of the Bond number.
\end{abstract}

\maketitle

\section{Introduction}
Loose granular packings, metastable granular structures and fragile granular
networks play an important role in a wide range of scientific disciplines,
such as collapsing soils \cite{Mitchell05,Barden1973,Assallay1997,Reznik05}, fine powders \cite{Rognon06} or complex fluids
\cite{wagner2009,delgado2010}. In collapsing soils without any doubt there is a metastable or
fragile granular network involved
\cite{Mitchell05,Barden1973,Assallay1997,kadau2009a,kadau2009c}. A similar
failure behavior can be found in colloidal gels \cite{Manley05} and snow
\cite{wu2005,Heierli05}.  But also powders have in
most cases an effective cohesive force, e.g.\ due to a capillary bridge
between the particles or van der Waals forces (important when going to very
small grains, e.g.\ nano-particles) leading to the formation of loose and
fragile granular packings \cite{Rognon06,kadau2009a,kadau2009b,kadau2009c}. In many complex fluids a
fragile/metastable network of colloids/grains is believed to be the essential
ingredient for the occurrence of shear thickening \cite{wagner2009} or yield stress behavior
\cite{delgado2010}.

The general feature of such fragile networks is that they can collapse/compact
under the effect of an applied load \cite{kadau2009b,kadau2009c,wu2005}. This load
can be an external load or exerted internally by a force acting on all particles
within the structure. This ``internal
collapse'' is important in different applications like cake formation of
filter deposits \cite{buerger2001,aguiar1996,stamatakis1990}, where the
compaction force in most situations is the drag force exerted on the grains by
the flow which is typically porosity dependent \cite{lu1995}. The structure's own
weight leads to compaction of snow after deposition \cite{ling1985} and during
aging \cite{Kaempfer2007,Vetter2010},  or to sediment compaction
\cite{bahr2001,bayer1989,athy1930}. In all cases, typically a depth dependent
porosity is observed and quantified by continuum descriptions
\cite{aguiar1996,stamatakis1990,ling1985,bahr2001,bayer1989,athy1930}.
In most cases the details of the porosity profile are  influenced by a
combination of 
different mechanical and chemical processes \cite{bahr2001,aguiar1996}. It is
well known  
that the porosity of a structure is of major importance for its mechanical
properties \cite{wu2005,bahr2001,roeck2008a,roeck2008b}, in filtration
processes \cite{araujo2006} and its chemical properties like
  catalytic activity
  \cite{kennedy2003}. 
The aim of this paper is to study the microscopic processes, i.e.\ on the grain
scale, for these internal compaction processes. For this,  we will investigate
the compaction due to gravity in a simplified model system of grains held
together by cohesive bonds. We analyze how the density profiles depend on the
granular Bond number, i.e.\  the ratio of cohesive force by gravity, and what
the influence of the dynamics of deposition/collapse.
\rev{
As discussed above loose structures are generated in nature, industrial application, experiments or
simulation by different processes. 
Here, we focus on ballistic deposition.
However, we expect the findings of this paper to be of relevance to all
systems involving compaction due to the particles' own weight. 

After a description of the simulation model and a brief discussion of possible
 experimental realizations in section \ref{sec:model}, we first study
the resulting density profiles when gravity acts during deposition, in
particular  the
influence of the granular Bond number (sec.\ \ref{sec:depo_grav}). To
understand the shape of the density profiles we study in the following
(sec.\ \ref{sec:slides})  the
role of the dynamics of the collapse occurring in small avalanches. We study
the average ``avalanche profile'' defined here as the average distance a
particle moves downwards after being deposited depending on its height. 
We observe characteristic profiles which can be used to relate the
 final density profile to the deposition density, given by the number of
 deposited particles per unit  volume (sec.\ \ref{sec:theoretical}). 
 To understand this phenomenological profile we study a simpler system where
 first all particles are deposited followed by the collapse of the whole
 structure leading to an even
simpler profile  (sec.\ \ref{sec:immediate}).  Our calculations yield that
this linear profile is 
obtained in all processes where a  homogeneous initial configuration is
collapsed/compacted  to a homogeneous final state.
In Sec.\ref{sec:relation} we show that the phenomenological obtained avalanche
profile obtained in sec.\ \ref{sec:depo_grav} can be derived from the linear
avalanche profiles of the homogeneous collapse.


}


 

\section{Description of simulation model} \label{sec:model}

The dynamical behavior of the system during generation is modeled with a
particle based method. Here we use a two dimensional variant of
contact dynamics, originally developed to model compact and dry systems
with lasting contacts \cite{Moreau94,Jean92,unger2003,brendel2004}.
The absence of cohesion between particles can only be justified in dry
systems on scales where the cohesive force is weak compared to the
gravitational force on the particle, i.e.\ for dry sand and coarser
materials, which can lead to densities close to that of random dense
packings. However, an attractive force plays  an important role in the stabilization of large voids
\cite{Kadau03}, leading to highly porous systems as e.g.\ in fine
cohesive powders, in particular when going to very small grain
diameters. 
 Also for contact dynamics a few simple models
for cohesive particles are established
\cite{taboada2006,richefeu2007,kadau2002,Kadau03}. Here we consider the bonding between two particles  in terms of a cohesion model with a constant attractive
force $F_c$ acting within a finite range $d_c$, so that for the
opening of a contact a finite energy barrier $F_cd_c$ must be
overcome. In addition, we implement Coulomb and rolling friction between two
particles in contact, so that large pores can be stable 
\cite{Kadau03,kadau_brendel_2003,bartels_unger_2005,brendel_kadau_2003,morgeneyer_roeck_2006}.

To generate the loose structure we use  \rev{ ballistic deposition where each
 deposited particle, chosen at random horizontal position, is attached to the
 structure at maximal possible height with zero velocity. 
At the same time we allow }   for all particles to move which can lead to a
partial collapse of the structures due to gravity
\cite{kadau2009a,kadau2009b,kadau2009c,kadau2010a}.
\rev{The structure is deposited on a flat surface, i.e.\ a wall at the bottom.
We use periodic boundaries in horizontal direction to avoid effects of side
walls, like Janssen effect.}
 During this process the
time interval between successive depositions crucially determines the structure
and density profiles of the final configurations.
 Here we will focus on the two extreme cases of very large time
intervals, i.e.\ the system can fully relax after each deposition of a single
grain, and vanishing time interval, i.e.\ the collapse of the systems
happens after the deposition process is complete. In the first case 
 the interval is chosen large enough to let the system compactify 
 and relax due
to the additional weight of the deposited grain. 
This is verified on the one hand by checking that the final
density is independent on the time interval and on the other hand by
monitoring 
the dynamics of the process. Having no time between depositions in practice means
that first we perform 
pure ballistic
deposition \cite{meakin86,meakin91}, and then switching on the full particle dynamics leading to a collapse
of the system due to
gravity. Experimentally, the two cases can be realized in a Hele-Shaw cell \cite{voeltz2000,voeltz2001,Vinningland07}
which can be tilted to effectively change gravity 
 In the slow
deposition process, simply the cell is slowly filled in an upright position
so that full gravity acts on the grains. In the other 
 case the
Hele-Shaw cell will be almost horizontal, so that the grains can be filled in
with nearly vanishing gravity, 
 and then the cell
is 
 tilted so that gravity can fully act on the grains,
leading to an abrupt collapse of the structure.
 
\section{Density profiles when gravity acts during deposition} \label{sec:depo_grav}

In this section we analyze the density profiles for the case of large enough
time intervals between successive depositions to allow the systems to relax
under the effect of gravity as described in the previous section. It is expected
that the density and the characteristics of the density profiles are mainly 
determined by the ratio of the cohesive force $F_c$ to gravity $F_g$, typically defined as
the granular Bond number $Bo_g=F_c/F_g$ \cite{rognon2006,nase2001}. Obviously
the case of $Bo_g=0$ corresponds to the cohesionless case whereas for
$Bo_g\to\infty$ gravity is negligible.  A similar
dimensionless quantity had been identified as most important parameter in
previous studies  on compaction of cohesive powders
\cite{Kadau03,kadau_brendel_2003,Wolf2005}.   

\begin{figure*}[htb]
\begin{center} 
\includegraphics[width=1.9\columnwidth]{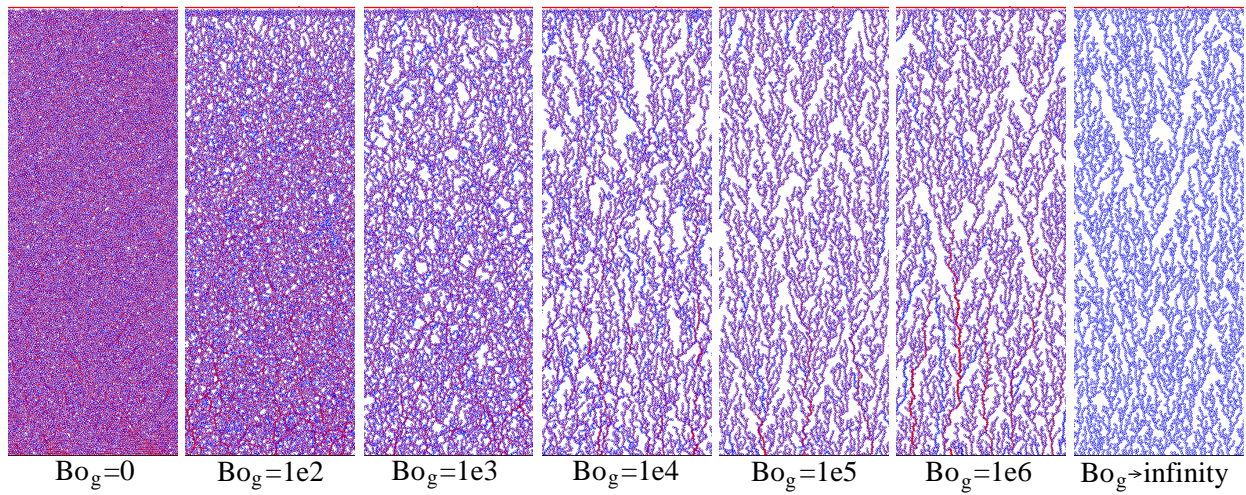}
\end{center}
\caption{\label{fig:structures} (Color online) Final structures achieved by the
  deposition/collapse process for different granular Bond numbers
  $Bo_g$. \rev{ In 
  addition to the particles compressive forces are illustrated by red (dark
  gray)  lines
  connecting the center of masses between the particles. In the case of
  $Bo_g\to \infty$ no forces are present as it is realized in the simulations
  by switching off gravity.}  }
\end{figure*}
In the following, we use monodisperse systems with a friction
coefficient $\mu=0.3$ and a rolling friction coefficient of $\mu_r=0.1$ (in
units of particle radii). The effect of varying these parameters is also
studied exemplary and will be discussed later. Typically the values
of the density can depend on these parameters as shown in
Ref.\ \cite{bartels_unger_2005} whereas the qualitative behavior does not change. 
Figure \ref{fig:structures} shows the final structures obtained for different
values of granular Bond number ranging from $0$ to $10^6$. Also the limit of
infinite Bond number is shown, leading to pure ballistic deposition
\cite{meakin91} well studied already in the past. For small Bond numbers, here represented by
$Bo_g=0$, the system typically  reaches a random close packing which also has been
studied intensively in the past. Note that our case of 
monodisperse particles typically leads in dense packings
to
crystallization effects which could be avoided by using a small
polydispersity. As our focus in this paper is on the looser structures where
this effect is not very important we prefer the monodisperse system to keep
the model as simple as possible.
In  the intermediate range of Bond numbers the density varies between the two limiting
values. 

\begin{figure}[htb]
\begin{center} 
\includegraphics[width=0.99\columnwidth]{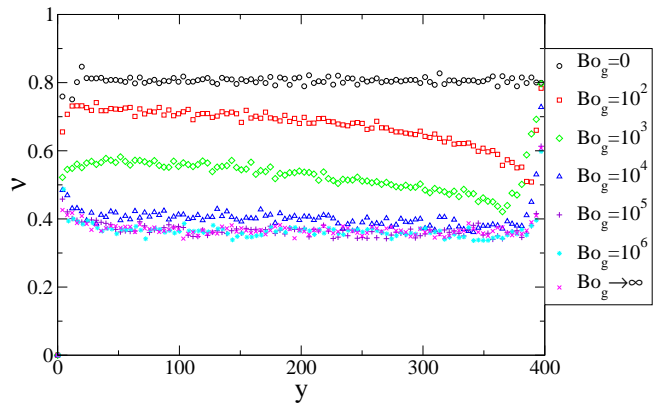}
\end{center}
\caption{\label{fig:density_profiles} (Color online) Density profiles for different granular
  Bond numbers $Bo_g$ (cf.\ fig.\ \ref{fig:structures}). Here, the volume
  fraction $\nu$ is plotted. 
  In this case the volume fraction is measured in thin slices
  of given width (here: $3.97$ particle radii) at varying height $y$.  For
  $Bo_g=0$ no cohesion is active and the random close packing is reached. 
  In the limit
  $Bo_g\to\infty$ the system does not collapse at all, and the simple ballistic
  deposition case \cite{meakin91} is obtained. 
}
\end{figure}
Plotting the density profile depending on the vertical position $y$
(Fig.\ \ref{fig:density_profiles})  provides a more quantitative analysis. It can
be seen that in the two limiting cases ($Bo_g=0$ and $Bo_g\to\infty$) the
density  \rev{is constant. For the infinite Bond number this can be explained
  easily as no collapse at all occurs and the density profile is that of a
  ballistic deposition and thus constant \cite{meakin86,meakin91}. For the
  non-cohesive case a close packing is expected, also leading to a constant
  density. This will be discussed again in more detail later in this paper.}
In the intermediate range the 
density decreases with increasing height. This is a result of the generation
process where the fragile structure is partially collapsed due to the weight
of the added particles which happens discontinuously in relatively small
avalanches as will be discussed in more detail in the next
section (sec.\ \ref{sec:slides}).

\begin{figure}[htb]
\begin{center} 
\includegraphics[width=0.9\columnwidth]{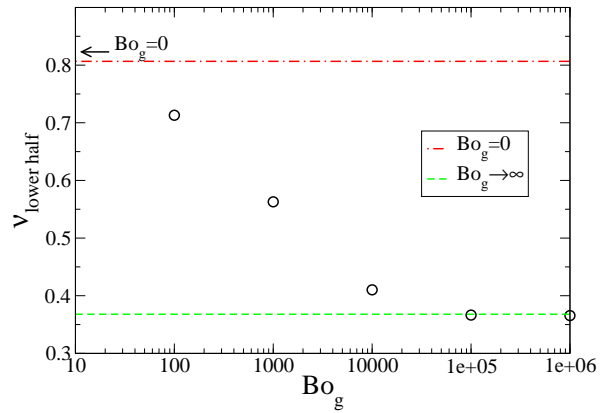}
\end{center}
\caption{\label{fig:density_Bog} (Color online) Average \rev{ volume fraction $\nu_{\rm
      lower\, half}$ depending}  on
  granular Bond number $Bo_g$. The density is averaged in the lower half of
  the system excluding the region very close to the bottom  to avoid border
  effects  \rev{ (here we excluded the region below the height of 50 particle
    radii so that clearly boundary effects are removed for all curves,
    cf.\ fig.\ \ref{fig:density_profiles}).}  The volume fractions vary between the two limits given by random
  close packing ($Bo_g=0$) and pure ballistic deposition ($Bo_g\to\infty$).  }
\end{figure}
Knowing that the density depends on vertical position a general dependence of
 the total density on the Bond number cannot easily be defined. Instead, for a
 given system size as in Fig.\ \ref{fig:density_profiles} the density at a
 fixed position 
 can be
 measured. In Fig.\ \ref{fig:density_Bog} the averaged density in the lower
 half excluding the region very close to the bottom is shown versus the
 granular Bond number. The density varies between the two limiting cases
 $Bo_g=0$ and $Bo_g\to\infty$. Note that the Bond number is plotted in a
 logarithmic scale, i.e.\ to see substantial changes of volume fraction the
 cohesive force or the gravitational force have to be changed by orders of
 magnitude. Particles with similar gravity and cohesive force will show the
 same typical behavior. As typically both forces depend on the size
 of the particles it appears to be natural to characterize the behavior of
 granular matter and powders by the grain size. 
 \revb{ For non-cohesive material recent experimental, numerical and
   theoretical studies \cite{shundyak2007,silbert2010,farell2010,wang2010} investigate the influence of the friction coefficient
   on, e.g.\ the volume fraction.}
  \rev{ \revb{A similar behavior as found here for the cohesive material when varying the
    granular bond number, has
    been found 
   \cite{shundyak2007,silbert2010}:}  varying the friction
   coefficient on a logarithmic scale  leads to a variation between the values
   $0.84$ for the packing fraction of a random close packing 
   and the value $0.77$ for infinitely large friction coefficient (in two
   dimensions, in three dimensions between $0.64$ and $0.55$). In the cohesive case as discussed here this range of
   accessible volume fractions is much higher and limited by the preparation
   protocol, i.e.\ in this paper by the ballistic deposition. This limit
  of course can be changed when changing the preparation protocol, e.g.\  by
  introducing a capture radius (cf.\ sec.\ \ref{sec:immediate}). }

\begin{figure}[htb]
\begin{center} 
\includegraphics[width=0.9\columnwidth]{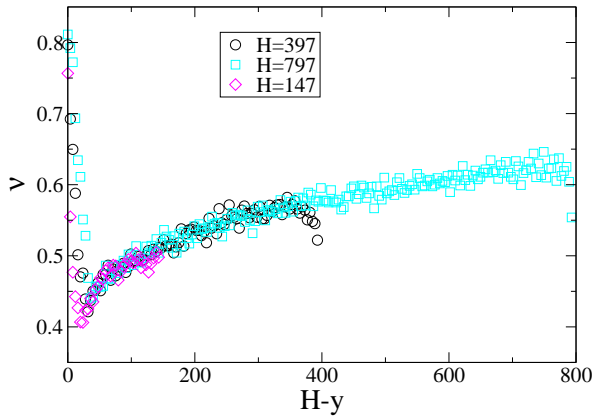}
\end{center}
\caption{\label{fig:density_size} (Color online) Illustration of the effect of 
  system size for intermediate density range (Bond number
  $Bo_g=10^3$). Plotting the depth $H-y$ measured from the surface $H$ of the
  final 
  packings smaller systems show the same profile as  large systems.  }
\end{figure}
For all results presented above  the total system height $H$ was fixed,
i.e.\ the deposition process stops when no more particles can be deposited
below a specified value $H$. When comparing density profiles for different
system heights $H$ plots depending on the vertical position $y$ will 
show different densities. A scaling can be achieved when plotting the density
versus the depth $H-y$ as illustrated in Fig.\ \ref{fig:density_size}. This
means the upper part of the large system is depositing and collapsing in the
same way as the small system while additionally leading to a further collapse
of the structure deposited previously below, accompanied by a downwards
motion of the whole upper part. Obviously the slow deposition process
guarantees that inertia is not important
 (cf.\ sec.\ \ref{sec:immediate}).  

\rev{ The specific behavior of the density profiles shown in this section 
  results from a deposition process combined with a collapse of the current
  structure due to gravity. The deposition is characterized by the
  number of deposited particles per volume, which we call ``deposition
  density'' and  which here is not constant (sec.\ \ref{sec:theoretical}).
  The collapse happens successively in relatively
  small avalanches, 
  analyzed in detail in the following section. In section
  \ref{sec:theoretical} we will show that these avalanches can be used to
  relate the final density profile to the ``deposition density''. }



\section{Analysis of the avalanches during deposition/collapsing} \label{sec:slides}

\begin{figure}[htb]
\begin{center}  
\includegraphics[width=0.99\columnwidth]{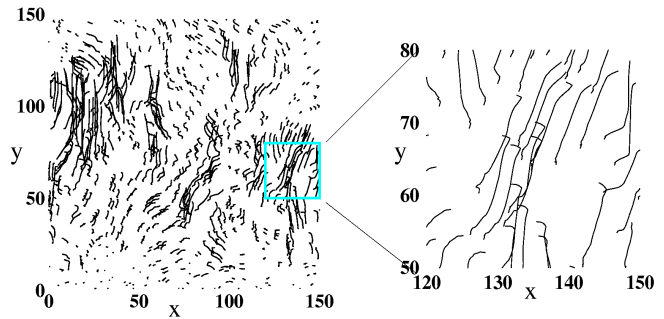}
\end{center}
\caption{\label{fig:trajects} Particle trajectories of particles of the
  small system   ($H=147$, cf.\ Fig.\ \ref{fig:density_size}) for the whole
  deposition/collapse. For better visibility only each 5th particle's
  trajectory  is  \rev{shown, i.e.\ the trajectories of 640 particles (instead
    of all 3200   particles).}  Viewing the total system (and on the right more
  detailed when  
  zooming in) illustrates that parts of the system
move collectively downwards accompanied by a sidewards motion or
rotation. When zooming in the individual trajectories can be identified which
are composed of the sum of paths 
during all the small avalanches
experienced by  the particle. } 
\end{figure}
Typically the collapsing of the structures, as mentioned earlier, happens
discontinuously in small avalanches. 
\rev{ As these avalanches are important also  for the final density profiles
  (see sec.\ \ref{sec:depo_grav}) their characteristics is studied in detail
  in this section.} 
 To illustrate the nature of these
avalanches in Fig.\ \ref{fig:trajects} the trajectories of the particles
are plotted for a relatively small system of height $H=147$ consisting of
about 3200 particles \rev{(for better visibility only each 5th trajectory is
  shown, i.e.\ the trajectories of 640 particles, instead of all 3200
  particles).}  
The avalanches are a collective motion of parts of the system. 
This mainly downwards motion is accompanied by a
sidewards motion 
or rotation. 
 When zooming in individual
trajectories can be identified. These trajectories represent the motion of
each particle during deposition/collapsing. Thus, they show the paths
that a particle experiences in all avalanches at different
times. 
Neighboring particles can have very similar trajectories, i.e.\ they belong to
the same set of avalanches at different times. 


\begin{figure}[htb]
\begin{center} 
\includegraphics[width=0.9\columnwidth]{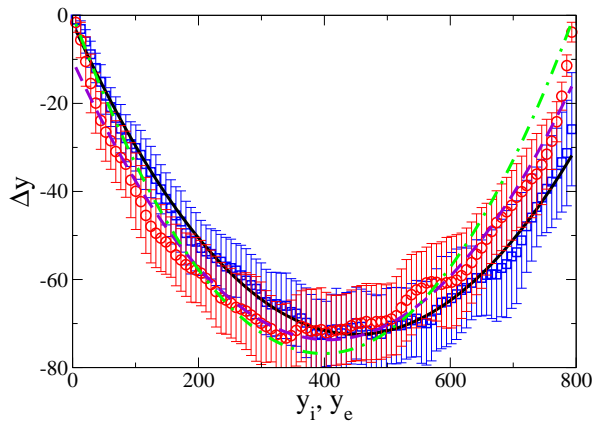}
\end{center}
\caption{\label{fig:dy_Bog1e3} (Color online) Size of avalanches depending on vertical
  position for $Bo_g=10^3$.  Here
  the size is measured by $\Delta y$, the average total downwards motion of a
  particle after deposition (initial position minus final position).  On the
  horizontal axis the initial position $y_i$ (blue squares) and final position
  $y_e$   (red circles) 
  are plotted. This leads to two slightly shifted curves as $y_e<y_i$.   The
  solid lines 
  represent parabolic fits $\Delta y (y_i)=-1.8-0.31y_i+0.00035y_i^2$ (black,
  full line)
  and  $\Delta y (y_e)=-10.2-0.31y_e+0.00038y_e^2$ (violet, dashed).
 Additionally shown is a fit by $\Delta y (y_e)=-a y_e(1-y_e/H)$ predicted by
 the considerations in sec.\ \ref{sec:relation} leading to $a\approx 0.39$
 (green, dashed-dotted). 
   }
%
\end{figure}
In Figure \ref{fig:dy_Bog1e3} we show the size of avalanches depending on initial and final vertical
position. This size is measured  by $\Delta y$, the total
downwards displacement of the particle after its deposition, i.e.\ initial position
minus final position. This represents for each particle the sum of all
avalanches occurring during 
the generation process, resulting in as many data points as
particles in the system. In Fig.\ \ref{fig:dy_Bog1e3} this data is averaged in bins of size
two particle diameters. The fluctuations within each bin are shown by the
vertical error bars. Both curves (for $y_i$ and $y_e$) can be relatively well
approximated by parabolas:
\begin{eqnarray}\label{eq:squarefit}
  \Delta y (y_i)=a'+b'y_i+c'y_i^2, \quad \Delta y
(y_e)=a+by_e+cy_e^2
\end{eqnarray}
 It is obvious that both curves cannot obey exactly the
parabolic behavior as $y_i$ and $y_e$ are related by $y_e(y_i)=y_i+\Delta
y(y_i)$. However, in the cases presented in this section, obtained by slow
deposition, the value of $\Delta y$ is relatively small compared to $y_i$ so
that $y_e(y_i)$ is very close to a straight line, leading only to a very small
horizontal shift. 
This behavior is typical for intermediate Bond numbers whereas in the limiting
cases no noticeable dependence of $\Delta y$ on the vertical position could be found. For $Bo_g=0$ a
 small constant value, below the particle diameter (around $1.5$ particle
 radii) is observed. In the case $Bo_g\to\infty$ no collapse happens, i.e.\ all $\Delta
 y=0$.

\begin{figure}[htb]
\begin{center} 
\includegraphics[width=0.9\columnwidth]{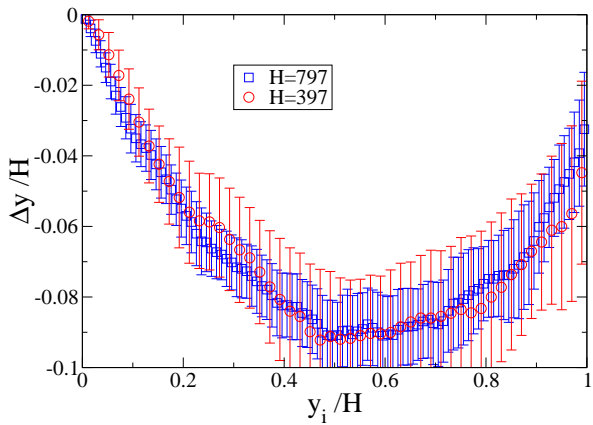}
\end{center}
\caption{\label{fig:dy_Bog1e3_scaled} (Color online) Collapse of the size of the avalanches
  for two different system sizes
  can be obtained scaling both axes by the system height (here: $Bo_g=10^3$). Under the
  assumption of a parabolic profile this scaling leads to a $1/L$ dependence
  of $c$ (pre-factor of quadratic term in  eq.\ \ref{eq:squarefit}).  }
\end{figure}
The parabolic behavior can be reproduced also for other system heights. In
Fig.\ \ref{fig:dy_Bog1e3_scaled} two different system sizes again for
$Bo_g=10^3$ are shown collapsed by scaling  both axes by the system
height $H$. From this scaling one can deduce the system size dependence of the
pre-factor of the quadratic term in eq.\ (\ref{eq:squarefit}). The scaling becomes:
\begin{equation}
  \label{eq:scaling}
 \Delta  y(y,H)= H\cdot f(y/H) \propto H\cdot (y/H)^2\propto 1/H
\end{equation}
when assuming that $\Delta y\propto y^2$ (parabolic behavior, see eq.\ \ref{eq:squarefit}). This $1/H$
dependence could be verified by fitting the curves in
Fig.\ \ref{fig:dy_Bog1e3_scaled}. 
Note that the parabolic shape was also found when varying the friction
coefficient $\mu$ and the rolling friction coefficient $\mu_r$. 

\begin{figure}[htb]
\begin{center} 
\includegraphics[width=0.9\columnwidth]{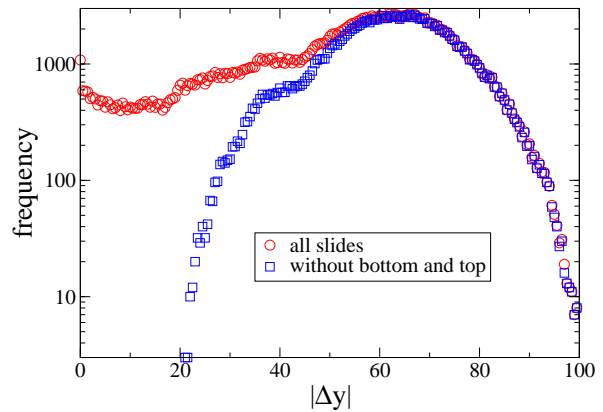}
\end{center}
\caption{\label{fig:histogram_dy_1e3} (Color online) The histogram of the \rev{size of
    avalanches $|\Delta y|$ for} 
  $Bo_g=10^3$ follows basically a Gaussian. Deviation from this behavior can
  almost fully be suppressed when removing the bottom and top part of the
  system.    }
\end{figure}
Whereas the average of the avalanche size $\Delta y$ as function of the
vertical position shows a parabolic profile 
of reasonable quality, there are of course large fluctuations around this
value. 
 In Fig.\ \ref{fig:histogram_dy_1e3}  we show
the distribution of the avalanche \rev{ sizes (here $|\Delta y|$)}  for the
entire system, i.e.\ independent 
on the vertical position. When removing the upper and lower part of the system
to decrease boundary effects we obtain a Gaussian distribution, i.e.\ we get
an estimate of a typical avalanche size. This typical size decreases with
increasing Bond number, and in the limit of $Bo_g\to\infty$, where no avalanches
occur, it vanishes.
\begin{figure}[htb]
\begin{center} 
\includegraphics[width=0.9\columnwidth]{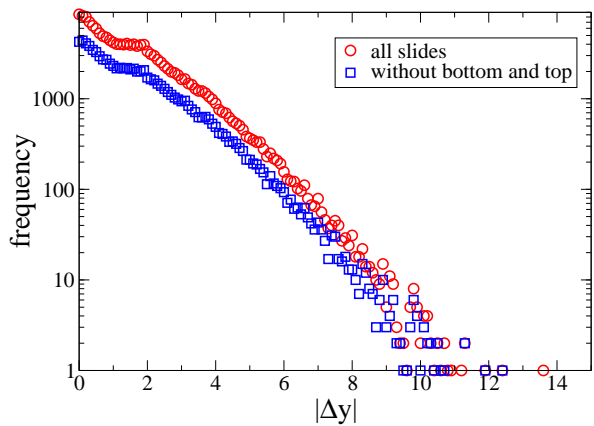}
\end{center}
\caption{\label{fig:histogram_dy_0} (Color online) The histogram of \rev{ avalanches $|\Delta
    y|$ for}  
  $Bo_g=0$ basically shows an exponential decay. Deviation from this behavior
  can be found \rev{  for $|\Delta y|$}  between 1 and 2 particle
  radii where the probability is about constant. This effect cannot be
  suppressed when removing the bottom and top   part of the   system (as,
  e.g.\  for $Bo_g=10^3$).    } 
\end{figure}
In the limit of $Bo_g=0$ (no cohesion) the behavior is different, an
exponential decay is obtained (Fig.\ \ref{fig:histogram_dy_0}). Here the
boundaries have no effect,  i.e.\ we get the same behavior when
removing  the upper and lower part of the system as done previously. For this
Bond number typically the surface of the structure during deposition grows
relatively flat,  so that \rev{ large $|\Delta
y|$ }  are unlikely as expressed by the exponential decay. Due to the
monodispersity this surface is locally almost a flat crystalline surface with
heaps 
which consists of a few particles only, in most cases one particle. When a
particle is deposited on a one particle heap it rolls off to rest eventually
as a ``crystalline'' neighbor aside the particle, resulting \rev{ in $|\Delta y|$}
between one and two. This leads to the very small range \rev{ of $|\Delta y|$}  with
constant probability in
Fig.\ \ref{fig:histogram_dy_0}. Taking a slightly polydisperse system this
region would disappear. 

\rev{ In this section we studied the collapse of the structures occurring in
  small avalanches we  analyzed statistically.   We suggest to
  characterize these avalanches by their ``size'', showing a
  typical dependence on vertical position, a parabolic shape for the
  specific systems investigated in this section. In the
  following section we will use this characteristic behavior to be able to
  derive the final density profile from the ``deposition density''. In section
  \ref{sec:immediate} the same concept will be shown to be applicable also for
  other protocols of generating loose structures. }

\section{Theoretical analysis of the avalanches} \label{sec:theoretical}

In the previous \rev{sections} we mentioned that the dynamics leading to a
final configuration is determined by small avalanches occurring during the
deposition process. All these compaction events 
contained in the function $\Delta y(y_e)$, which is given by the difference
between the initial position $y_i$ and final position $y_e$. Note that $\Delta
y$ can be plotted (e.g.\ fig. \ref{fig:dy_Bog1e3}) as function of the final
position $y_e$ or alternatively as function of the position of deposition
$y_i$. The aim of this section is to relate the final density profile to the
dynamic process of deposition and collapse by using $\Delta y(y_e)$, showing
how the avalanches produce the final density $\rho_f(y_e)$ from the deposition
density $\rho_d(y_i)$. The deposition density is defined by the number of
particles deposited within a volume.  As the structure collapses between the
depositions the deposition density is not independent on the collapsing, and
it is possible that at (almost) the same position several particles are
deposited. Thus, locally within a fixed volume even more particles could be
deposited than typical for a dense packing.
 

We first calculate the number $N_{d,f}$ of particles
up to a given height $y_{i,e}$ ($L_x$ width of the two dimensional system in
units of particle radii):
\begin{equation}
  \label{eq:Number}
 N_{d,f}(y_{i,e}) = L_x \int_0^{y_{i,e}} dy' \rho_{d,f}(y')
\end{equation}
The final position $y_e$ of particles can be related to the position $y_i$ of
deposition by the avalanche profile $\Delta y$:
\begin{equation}
  \label{eq:ye_relation} 
  y_i(y_e)=y_e - \Delta y (y_e), \qquad\mbox{or }\quad  y_e(y_i)=y_i + \Delta y (y_i)
\end{equation}
In this notation $\Delta y$ is negative as the motion of the particles is
downwards  (due to gravity). Therefore, $y_i$ \rev{is larger than or equal to
  $y_e$.} 
As particles are never destroyed the number of particles deposited up to a
given height, $N_d(y_i)$, will stay the same, but shifted to a lower height,
$N_f(y_e)$, where $y_i$ and $y_e$ are related by eq.\ 
(\ref{eq:ye_relation}). Together with eq.\ (\ref{eq:Number}) this leads to:
\begin{equation}
  \label{eq:NdNf}
 N_f(y_e)/L_x = N_d(y_i(y_e))/L_x =  \underbrace{\int_0^{y_i(y_e)} dy'
   \rho_d(y')}_{\equiv G(y_i(y_e))}
\end{equation}
This relates $N_f$ to the deposition density whereas eq.\ (\ref{eq:Number})
relates $N_f$ to the final density. The function $G$ here is
formally introduced for the integral as abbreviation, by derivation of $G$ the
density is retrieved. 
The final density can be obtained by derivation of $N_f/L_x$ using eq.\
 (\ref{eq:NdNf}):
\begin{eqnarray}
\rho_f(y_e) = \frac{d}{dy_e} \frac{N_f(y_e)}{L_x}
&=& \frac{d}{dy_e} G(y_i(y_e)) 
         = \frac{dG(y_i)}{dy_i} \frac{dy_i}{dy_e} \nonumber \\ 
    &=& \rho_d(y_i(y_e)) \frac{dy_i}{dy_e}  \label{eq:rhof_calc}\\ 
         &=& \rho_d(y_i(y_e)) \left(1- \frac{d\Delta y (y_e)}{dy_e}\right)
         \nonumber   
\end{eqnarray}
 The deposition density
$\rho_d(y_i(y_e))$ in principle can be expressed directly by $y_e$ introducing
$\rho_d'(y_e)$. As usually the functional behavior of both  functions
is not known, but only values for specific $y_i$ and $y_e$, the transformation
can be done for each point by simply using eq.\ (\ref{eq:ye_relation}),
i.e.\ replacing each $y_i$ by $y_e=y_i+\Delta y (y_i)$.  Summarizing, to
calculate the final density profile one needs to know the deposition density
$\rho_d$ and the avalanche profile $\Delta y$. Note that the avalanche profile
dependence on both $y_e$ and $y_i $ is needed, which can be calculated from
each other for
some cases as shown later. For experimental
situations these quantities are not known. However, the relation between
$\rho_f$ and $\rho_d$ (eq.\ \ref{eq:rhof_calc}) can be used to
calculate the deposition density from the final density in the
slow deposition limit, when assuming a parabolic profile as found in the
simulations before.

\begin{figure}[htb]
\begin{center} 
\includegraphics[width=0.9\columnwidth]{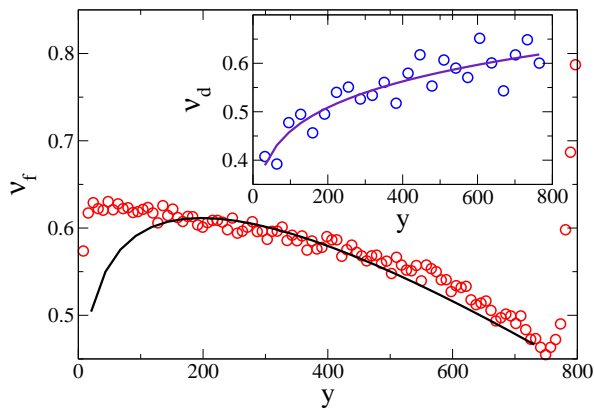}
\end{center}            
\caption{\label{fig:calculation_byslides1e3} (Color online) Using the parabolic approximation
  (fig.\ \ref{fig:dy_Bog1e3}) 
  for the average avalanches the final density (here volume fraction $\nu_f$) can be calculated from the deposition
  density (here shown: volume fraction $\nu_d$ for $Bo_g=10^3$, inset). There are strong fluctuations in the deposition density which are
  induced by the irregularity of the avalanches. To obtain a smooth curve we
  use a fit function (here: power law fit, with exponent $0.15$) to calculate
  the final density (black line) matching relatively well with the measured
  density profile for sufficiently large $y$ (except close to the bottom)
  }
\end{figure}
In figure \ref{fig:calculation_byslides1e3} we use eq.\ (\ref{eq:rhof_calc})  to
calculate the final density from the deposition density by using the parabolic
fit for $\Delta y$ (fig.\ \ref{fig:dy_Bog1e3}). In practice first the
deposition density curve is shifted on the horizontal axis by
$y_e=y_i-(a'+b'y_i+c'y_i^2)$, then multiplying the deposition density \rev{
  with the right hand side of eq.\ (\ref{eq:rhof_calc}),  $1-(b+2c y_e)$,
  i.e.\ using the derivative of $\Delta y(y_e)$ }  which is a linear
function. If the 
deposition density would be constant this would lead to a linear  profile for
the final density. However, the deposition density is not constant,
explaining the non-linear behavior for the final density. Additionally the
deposition density shows strong fluctuations, but 
by assuming the avalanches to follow the averaged parabolic behavior
the corresponding fluctuations in the avalanche profile are not included. 
The calculated curve matches relatively well the profile measured in the
simulations for sufficiently large values of the vertical position. Close to the
bottom, however, the calculated curve deviates from the measured one. In this
region the deposition density is very small, i.e.\ almost the one of pure
ballistic deposition. This can be understood as the system needs to gain a
sufficient amount of weight for the collapse to start (cf.\ also
sec.\ \ref{sec:immediate}). This should correspond to a higher initial slope
of $\Delta y (y_e)$ which is not reflected in the parabolic approximation
(eq.\ \ref{eq:squarefit}). In this region a higher order terms would be
necessary for reproducing also the system bottom.

\begin{figure}[htb]
\begin{center} 
\includegraphics[width=0.9\columnwidth]{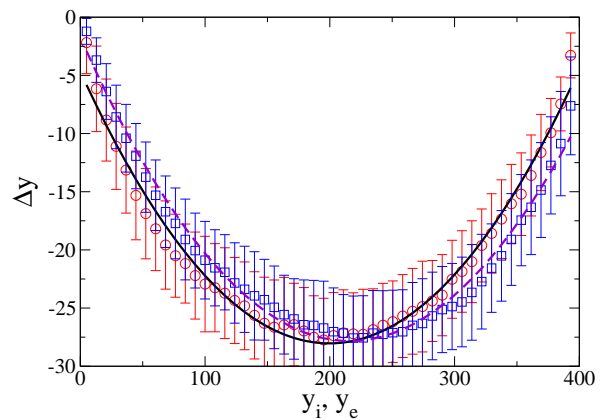}
\end{center}
\caption{\label{fig:dy_Bog1e2} (Color online) Size of avalanches depending on vertical
  position for $Bo_g=10^2$.  Here
  the size is measured by $\Delta y$, the total downwards motion of the
  particle after deposition (initial position minus final position). On the
  horizontal axis the initial position \revb{$y_i$}  \rev{(blue squares)}  and final position
  \revb{$y_e$}
 \rev{(red circles)  are plotted. The lines
  represent the parabolic fits $\Delta y (y_i)=-1.7-0.24y_i+0.00056y_i^2$
  (dashed, violet) and $\Delta y (y_e)=-4.7-0.23y_e+0.00059y_e^2$ (full line,
  black). }    }
\end{figure}
 \begin{figure}[htb]
\begin{center} 
\includegraphics[width=0.9\columnwidth]{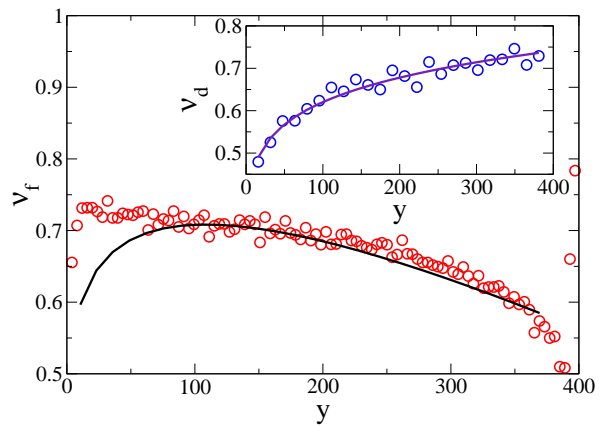}
\end{center}
\caption{\label{fig:calculation_byslides1e2} 
(Color online) Using the parabolic approximation
  (fig.\ \ref{fig:dy_Bog1e2})
  for the average avalanches the final density can be calculated from the deposition
  density (here shown: volume fraction for $Bo_g=10^2$). There are strong fluctuations in the deposition density which are
  induced by the irregularity of the avalanches. As before (for $Bo_g=10^3$),
  to obtain a smooth curve we use a fit function (here: power law fit, with
  exponent 0.13) to calculate the final density (black line) matching
  relatively well for sufficiently large $y$ (except close to the bottom).
}
\end{figure}
 The same analysis has been done also, e.g.\ for $Bo_g=10^2$ as shown in
 Figs.\ \ref{fig:dy_Bog1e2} and \ref{fig:calculation_byslides1e2}.  In this case
 the deposition density shows somewhat lower fluctuations as for $Bo_g=10^3$
 (Fig.\ \ref{fig:calculation_byslides1e3}). \rev{ To quantify this we 
   estimated the fluctuations of the deposition density at vertical position
   $y=200$ for both cases.  For  $Bo_g=10^2$ we obtained around 15\% whereas
   we estimated around 20\%  for $Bo_g=10^3$. 
For the case of \revb{$Bo_g\to\infty$} there is no avalanching at all
(cf.\ sec.\ \ref{sec:slides}), and trivially the final density equals the
deposition density. This is very similar for very large $Bo_g$, but as some
avalanches occur there are some relatively small fluctuations in the average
profile.  Away from this limit, but still close enough
that the density profile is very similar to the \revb{$Bo_g\to\infty$} case,  as
e.g.\  for $Bo_g=10^4$,  very large
fluctuations in the avalanche profile are observed. Thus, the parabolic
profile cannot easily be identified.  Still the
theory works well as the final density is very close to the deposition density
so that even a very inaccurate fit for the avalanche profile does not affect
the calculated density profile too much. In the limit of $Bo_g=0$ the
avalanche profile is a constant (cf.\ sec.\ \ref{sec:slides}), i.e.\  all
grains are slightly shifted downwards by the same amount (except boundary
effect at the bottom).  As then the
derivative vanishes the final density equals the deposition density. 
 } 

\rev{Here, we showed how the parabolic avalanche profile can be used to
  calculate the final density profiles from the deposition density in the case
where gravity acts during deposition. In the next section the same concept
will be used to the simpler case of collapse of the system after deposition is
complete. These two cases could be then related in section
\ref{sec:relation}. }
 
\section{Collapse after deposition complete} \label{sec:immediate}

\rev{In the previous sections we investigated the case where gravity acts
  during deposition, leading to relatively complex shape of the density
  profiles and a parabolic characteristics of the avalanche size. For this
  case we showed that these avalanche profiles can be used to relate the final
  density profile to the deposition density. }
In this section we will analyze the case when the particles are first
deposited, then gravity is switched on and the structures collapse. \rev{This
  case is even simpler and can later be used to understand the more complex
  system studied before.}  In this
case  the initial density $\rho_i$ characterizes the system (instead of the
deposition density as in  the previously discussed situation).
\begin{figure*}[htb]
\begin{center} 
\includegraphics[width=1.8\columnwidth]{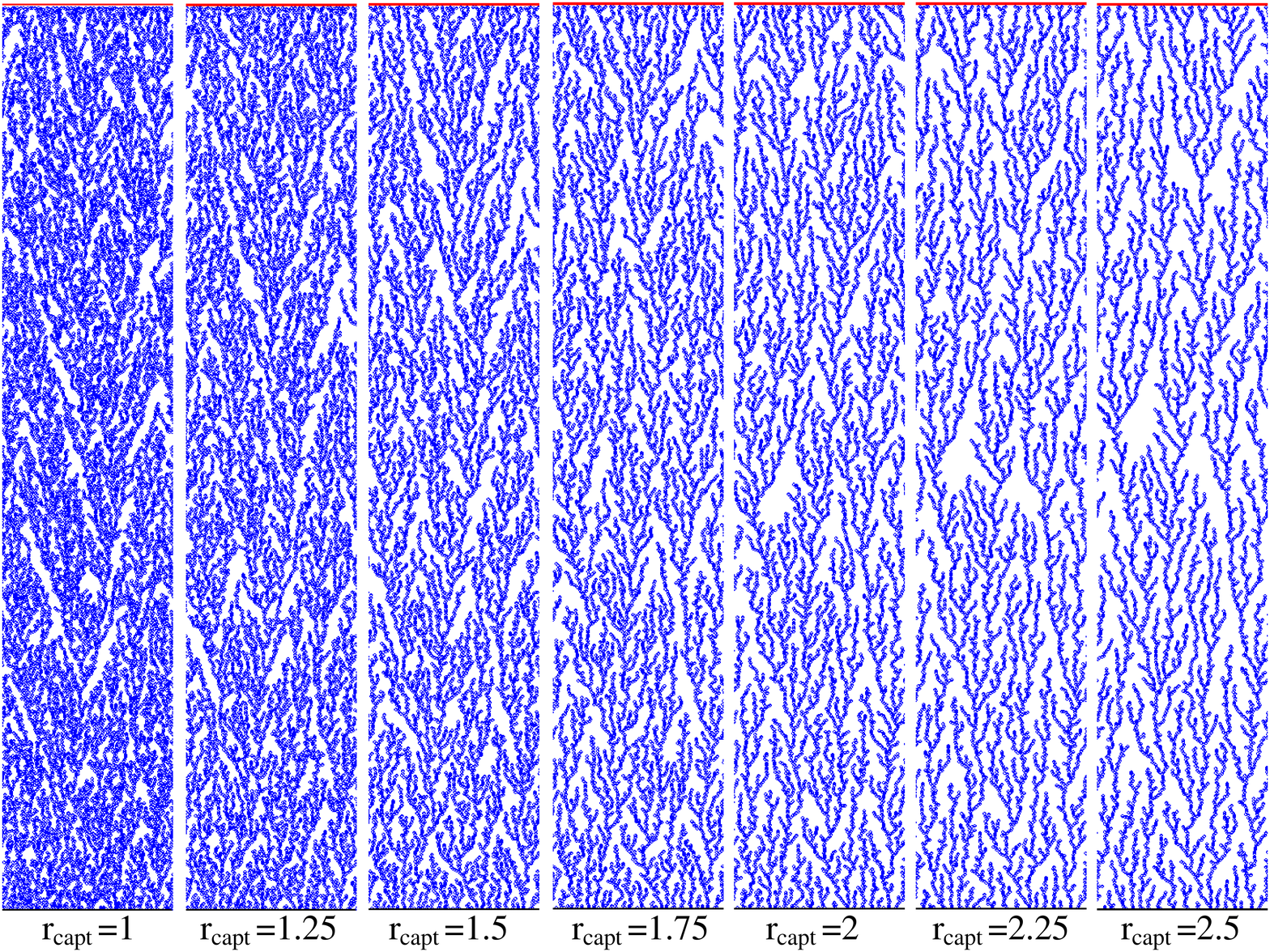}
\end{center}
\caption{\label{fig:initial_dens} Initial structures generated by ballistic
  deposition with increasing capture radius $r_{\rm capt}$.  }
\end{figure*}
Using the off lattice version of  ballistic deposition
as presented in Refs.\ \cite{meakin86,meakin91}  with sticking probability one, vertically
falling particles stick when they touch 
an already deposited particle. This leads to a fixed initial density. 
Lower densities can be obtained by using a capture radius $r_{\rm capt}$ i.e.\ particles
stick to each other when they are within a certain distance during the falling
of the depositing particle. 
 More precisely:
When the distance between the center of masses of two particles is below $2\cdot
r_{\rm capt}$ the particles stick and the falling particle is pulled along the
connecting line towards the already deposited particle. This capture radius is a
measure for the distance between the branches of the deposit and the resulting
density is inversely proportional to $r_{\rm capt}$ \cite{Kadau03}, $r_{\rm
  capt}=1$  gives  the original method.  The 
resulting initial structures are shown in Fig.\ \ref{fig:initial_dens}. These
structures obtained with different capture radius will be later used to study
the influence of the initial density.

\begin{figure}[htb]
\begin{center} 
\includegraphics[width=0.9\columnwidth]{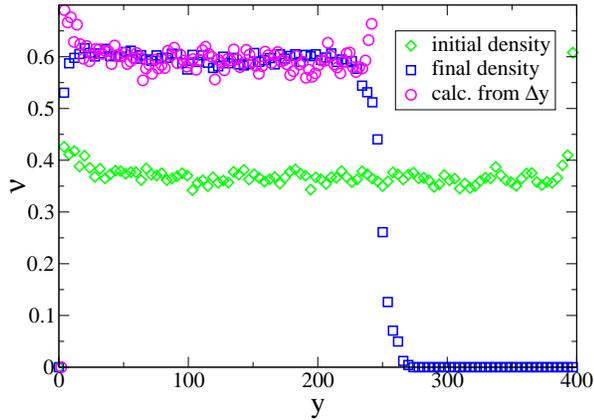}
\end{center}
\caption{\label{fig:immediate_dens} (Color online) The initial and final density are about
  constant when depositing first and then collapsing the system
  ($Bo_g=10^3$). Using the 
  linear dependence of the avalanches $\Delta y$ on the vertical position (see
  Fig.\ \ref{fig:immediate_dy}) the
  final density can be calculated from the initial density using eqs.\ 
  (\ref{eq:ye_relation}) and (\ref{eq:rhof_calc}). The results support the
  analytical considerations.  }
\end{figure}
 First we will investigate the behavior  using
 $r_{\rm capt}=1$. Figure \ref{fig:immediate_dens} shows the density profile
 before the collapse which is the same that we got in the limit of
 $Bo_g\to\infty$ in sec.\ \ref{sec:depo_grav}, also independent on vertical
 position. After this deposition is
 complete gravity is ``switched on'' and the structure abruptly collapses. Here
 we choose a Bond number of $Bo_g=10^3$. This leads  to a final structure with
 higher density, in this case also independent on the vertical position
 (Fig.\ \ref{fig:immediate_dens}). As no particles are added after the initial
 deposition the final system height is lower.  

\begin{figure}[htb]
\begin{center} 
\includegraphics[width=0.9\columnwidth]{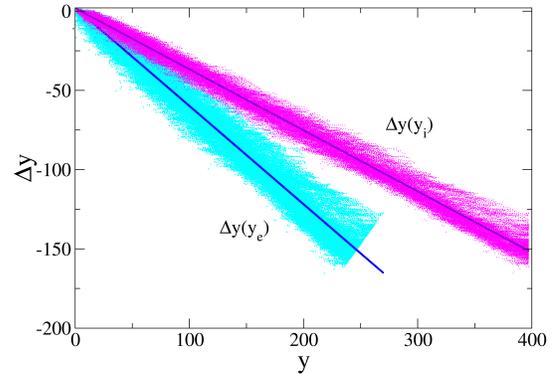}
\end{center}
\caption{\label{fig:immediate_dy} (Color online) The linear dependence of the avalanche sizes
  $\Delta y (y_{i,e})$ explains the homogeneous density increase as seen in
  fig.\ \ref{fig:immediate_dens}. The linear fits are $\Delta y
  (y_e)=2.2-0.62y_e$ and  $\Delta y(y_i)=1.9-0.39y_i$. 
 }
\end{figure}

Similarly as we did before we analyze 
 the size of the
avalanches $\Delta y$ as defined in sec.\ \ref{sec:slides}. 
Figure \ref{fig:immediate_dy} shows a linear dependence of  $\Delta y$  on
either $y_e$ and $y_i$. The fit parameters of the two lines can be related to
each other by the relation between $y_i$ and $y_e$
(eq.\ \ref{eq:ye_relation}). Assuming  $\Delta y (y_e)=a-by_e$ and $\Delta y
(y_i)=a'-b'y_i$ the values $a'$ and $b'$ can be calculated from $a$ and $b$
(see app.\ \ref{sec:relation_linear}) as:
\begin{equation}
  \label{eq:relation_linear}
b'=\frac{b}{1+b}, \quad a'=\frac{a}{1+b}
\end{equation}
The vertical dependence of  $\Delta y$ can be used similarly as before to
calculate the final density from the initial density by using
eq.\ (\ref{eq:rhof_calc}). The calculated density profile using this linear
dependence reproduces the obtained final density profile very well as shown in
Fig.\ \ref{fig:immediate_dens}. In this case the agreement is better
as now the initial density is not fluctuating very much in contrast to the
cases discussed in sec.\ \ref{sec:theoretical}. The density
increase $\Delta\rho$ (or volume fraction increase $\Delta\nu$) can be directly calculated by the constant slope of $\Delta y(y_e)$:
\begin{equation}
  \label{eq:deltanu}
\frac{\Delta \rho}{\rho_i}=\frac{\Delta \nu}{\nu_i} = - \frac{d\Delta y (y_e)}{dy_e}
\end{equation}

\begin{figure}[htb]
\begin{center} 
\includegraphics[width=0.9\columnwidth]{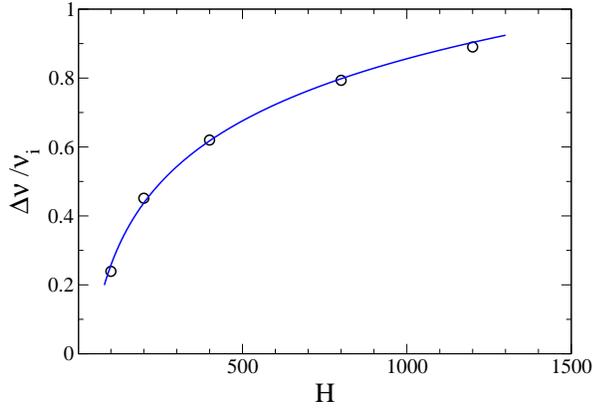}
\end{center}
\caption{\label{fig:slope_size} Dependence of volume fraction increase $\Delta
  \nu/\nu_i$ on  system height $H$ when
  first deposited and then collapsed. A logarithmic fit matches the data best
  (here $y=-0.94 + 0.26 \ln(x)$). The limit of random close packing defines the
  largest possible value for  $\Delta\nu/\nu_i$ of $1.19$, which will be
  approached for infinite system heights. 
} 
\end{figure}
For the same parameters ($Bo_g=10^3$) we studied the effect
of the  system height $H$
on the density  increase while still keeping the initial density fixed
(fig.\ \ref{fig:slope_size}). A logarithmic fit matches the data best. This fit certainly cannot continue to
  infinity as there is a limit for the density $\rho_{\rm max}$ given by the
  random close packing (see also 
   fig.\ \ref{fig:density_Bog}), leading to a $(\Delta\nu/\nu_i)_{\rm max}$ of
   $1.19 (\simeq \rho_{\rm
  max}/\rho_{\rm ini}-1)$. 

\begin{figure}[htb]
\begin{center} 
\includegraphics[width=0.9\columnwidth]{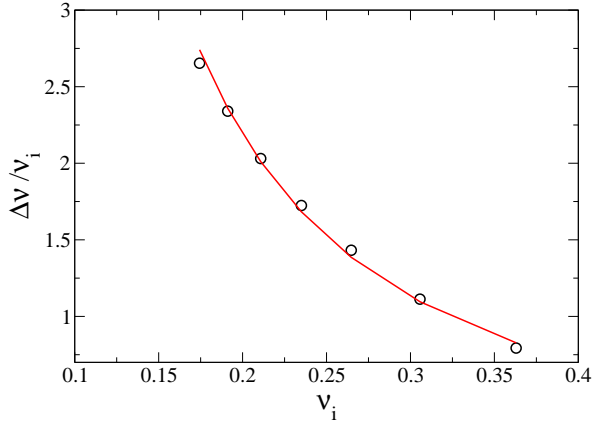}
\end{center}
\caption{\label{fig:slope_dens} Dependence of volume fraction increase $\Delta
  \nu/\nu_i$ on volume fraction $\nu_i$ of the initial
  system  (system first deposited and then collapsed). Different densities 
  could be reached by increasing the capture radius for ballistic deposition
  (fig.\ \ref{fig:initial_dens}). A power law fit with
 \revb{ \rev{ exponent $1.64$
}}   fits relatively well (power law fit results in $y=0.158 x^{-1.64}$). }
\end{figure}
Using initial capture radii as described above we study the influence of the
initial density on the relative density increase  $\Delta\nu/\nu_i$
(fig.\ \ref{fig:slope_dens}). We could obtain  the best fit when using a
power law with exponent of \revb{\rev{ about $1.64$.
}}

\rev{ We showed in this section that the linear avalanche profile is a
  characteristic feature of compacting from a depth independent to a depth
  independent structure, obtained here for systems generated by ballistic
  deposition collapsing due to gravity. More complex avalanche profiles with
  non-constant derivative will transform homogeneous structures into
  inhomogeneous structures. Thus, we expect the linear profile to be obtained
  in all cases where a homogeneous initial system compacts to a homogeneous
  final system. These homogeneous compaction processes are investigated in
  different research areas as e.g.\ discussed in
  Refs.\ \cite{kjeldsen2006,lumay2007,dong2006,valverde2006,son2008}. In
  addition in the next section we will show that also for the more complex
  process when gravity acts during deposition (sec.\ \ref{sec:depo_grav}) this
  linear profile can be used to derive the parabolic profile of the
  avalanches. }

\section{Relation between deposition under gravity and switching on gravity
  after deposition} \label{sec:relation}

For the very fast process a \revb{linear profile for $\Delta y$ depending on
  vertical position  has been found (cf.\ fig. \ref{fig:immediate_dy})}
whereas the slow deposition limit shows a \revb{\rev{ parabolic profile for $\Delta y$ depending on vertical position (cf.\ figs. \ref{fig:dy_Bog1e3},\ref{fig:dy_Bog1e2}).}}  In this section
we will discuss how a relation between both can be established. 
\rev{By this relation also the parabolic profile is put onto a more
  fundamental basis like the linear profile for the homogeneous collapse.}

\begin{figure}[htb]
\begin{center} 
\includegraphics[width=0.9\columnwidth]{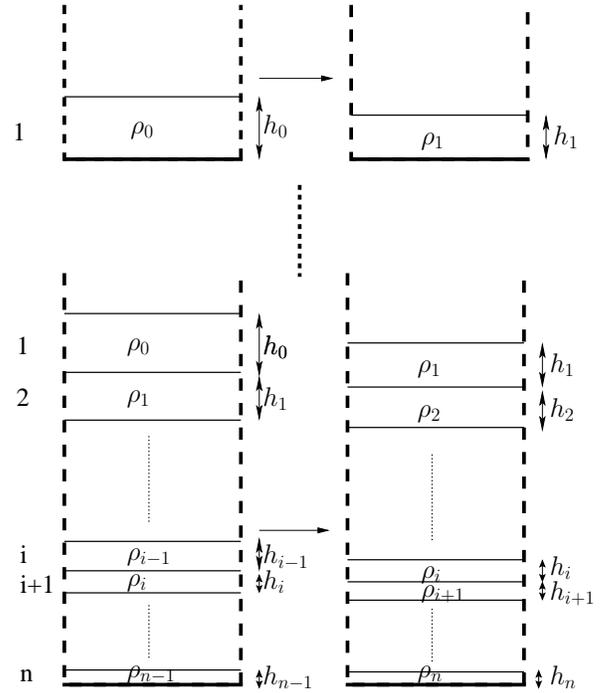}
\end{center}
\caption{\label{fig:sketch_slices} Sketch illustrating the procedure
of depositing the grains slice by slice. The first slice deposited is
compacted by internal collapse. The same is true for each ``freshly'' deposited
slice.
 The slices below are compacted by the added weight of the slices
above. Periodic boundary conditions in horizontal direction are imposed
(illustrated by dashed lines). The figure also illustrates the definition of
the symbols used here. The $n$ slices are numbered from $1$ to $n$. A slice $i$
collapses from $\rho_{i-1}$ to $\rho_i$ while its height decreases from
$h_{i-1}$ to $h_i$, where $h_i=\rho_{i-1}/\rho_i h_{i-1}$. }
\end{figure}

 Let us
imagine depositing particles slice by slice as sketched in Fig.\ \ref{fig:sketch_slices}.
The slices are thin parts of the system in vertical direction  spanning
the full system width in horizontal direction. They can be considered as
systems with 
very small initial height $h_0$. 
 In each slice the
deposition will be immediately followed by the collapse. However, there will
be not 
only an ``internal collapse'' within the ``freshly'' deposited slice, but also a
compaction of the slices below 
due to the additional weight of the ``freshly'' deposited slice.

Let us first consider systems composed of a small number $n$ of slices. The case
$n=1$ (one slice) is the same as discussed in the previous section: The system
collapses ``internally'' leading to an increase of the density from $\rho_0$
to $\rho_1$ while the height reduces from $h_0$  to $h_1$. Here we denote the
slice number as $1$ (cf.\ Fig.\ \ref{fig:sketch_slices}). As shown in the
previous section the avalanche sizes have
a linear profile $\Delta y(y_e^{(1)})=S_1y_e^{(1)}$. $S_1$ is the slope in slice
$1$, and is
the same for all freshly deposited slices when the height $h_0$ is kept
constant. The vertical position $y_e^{(1)}$ within slice $1$ is measured
from its bottom ($y_e^{(1)}=0\dots h_1$). This notation will be used in the
following for each slice $i$: $y_e^{(i)}=0\dots h_i$. 
The case $n=2$ (two slices) means adding an additional slice to the case
$n=1$. Then the lower slice (slice 2) experiences an additional compaction by the added
weight expressed by the corresponding avalanche size $C_2 y_e^{(2)}$ assuming
  a linear behavior for this relatively fast process similar as for the
  internal collapse. This is justified at least for the limit of small slices
  later considered in this section.  
  The
  upper slice (slice 1) will be compacted internally and additionally will move
  downwards by $C_2h_2$ ($=h_1-h_2$) as the slice below is compacted. Summarizing for the
  two slices we get:
  \begin{eqnarray}
    \label{two_slices}
    \Delta y^{(1)}(y_e^{(1)}) &=& S_1y_e^{(1)} + \underbrace{C_2 h_2}_{h_1-h_2} \\
    \Delta y^{(2)}(y_e^{(2)}) &=& C_2y_e^{(2)} + S_1
    \underbrace{(1+C_2)y_e^{(2)}}_{y_e^{(1)}}  \nonumber
  \end{eqnarray}
For slice 2 the internal compaction from the first step $S_1y_e^{(1)}$ has
been transformed by using that $\rho_2=(1+C_2)\rho_1$
(cf.\ eq.\ \ref{eq:deltanu}), leading to $h_2=\rho_1/\rho_2
h_1=1/(1+C_2)h_1$. 
Adding a further slice leads to the case $n=3$, where the two slices are
compacted due to the additional weight. Each of these compactions is
accompanied by a downwards shift of the slices above. This leads to:
  \begin{eqnarray}
    \label{three_slices}
    \Delta y^{(1)}(y_e^{(1)}) &=& S_1y_e^{(1)} + \underbrace{C_2 h_2+C_3
      h_3}_{(h_1-h_2)+(h_2-h_3)=h_1-h_3} \\
    \Delta y^{(2)}(y_e^{(2)}) &=& C_2y_e^{(2)}+
    \underbrace{S_1(1+C_2)y_e^{(2)} + C_2 h_2}_{\mbox{from step 2}}
    +C_3 h_3 \nonumber \\
    \Delta y^{(3)}(y_e^{(3)}) &=& C_3y_e^{(3)} \nonumber \\&& + \underbrace{C_2(1+C_3)y_e^{(3)}
    +S_1(1+C_2)(1+C_3)y_e^{(3)}}_{\mbox{from step 2}}  \nonumber
  \end{eqnarray}
Imagining continuing this iterative procedure, one obtains the case of $n$
slices. For the top slice this results in:
  \begin{eqnarray}
    \label{top_slice}
   \hspace{-3ex} \Delta y^{(1)}(y_e^{(1)}) &=& S_1y_e^{(1)} + \underbrace{C_2 h_2+C_3
      h_3+\dots +C_n h_n}_{h_1-h_n}
  \end{eqnarray}
The first term is the internal collapse where the other terms are the shift due
to the compaction of all slices below ($2$ to $n$) in this last step.
For the bottom slice we get:
  \begin{eqnarray}
    \label{bottom_slice}
\Delta y^{(n)}(y_e^{(n)}) &=& S_1(1+C_2)(1+C_3)\cdot\ldots\cdot(1+C_n)y_e^{(n)}
\nonumber \\
 && C_ny_e^{(n)} + C_{n-1}(1+C_n)y_e^{(n)}  +\dots  \\
 &&+ C_2(1+C_n)(1+C_{n-1})\cdot\ldots\cdot (1+C_3) y_e^{(n)} \nonumber
  \end{eqnarray}
Here all terms represent a collapse in the slice either internally by its own
weight when deposited in the first step, or when collapsing due to added
weight in the following steps. In addition these collapses have to be
transformed to a $y_e^{(n)}$ dependence (see above). 
For an arbitrary  slice $i$ somewhere in the system we get both types of terms
as in eqs.\ (\ref{top_slice}) and (\ref{bottom_slice}):
  \begin{eqnarray}
    \label{intermediate_slice}
\Delta y^{(i)}(y_e^{(i)}) &=&  S_1(1+C_2)\cdot\ldots\cdot(1+C_i)y_e^{(i)}
\nonumber\\ 
&& C_iy_e^{(i)} + C_{i-1}(1+C_i)y_e^{(i)} \\
&& +\dots +C_2(1+C_3) \cdot\ldots\cdot(1+C_i) y_e^{(i)} \nonumber\\
&& \underbrace{+C_nh_n+C_{n-1}h_{n-1}+\dots+C_{n-i+1}h_{n-i+1}}_{h_{n-i}-h_n}
\nonumber\\
&&\underbrace{+C_{n-1}h_{n-1}+\dots+C_{n-i}h_{n-i}}_{h_{n-i-1}-h_n}\nonumber\\
&& \vdots\nonumber\\
&&\underbrace{+C_{i+1}h_{i+1}+\dots+C_2h_2}_{h_1-h_{i+1}}\nonumber
  \end{eqnarray}
The part of the expression independent on $y_e^{(i)}$ represents the shift due
to compaction by the weight of the above added slices in $n-i$ steps
considering all slices below. It consists of $n-i$ times $i-1$ terms and  can
be written shortly as:
  \begin{eqnarray}
    \label{slice_simple}
\Delta y^{(i)}_{\rm shift}&=& \sum_{j=1}^{n-i} h_j-h_{j+i}
  \end{eqnarray}

The limit of large $n$ while keeping the total system height constant  gives
very small slices where the part $\Delta y^{(i)}_{\rm shift}$
dominates as for very small systems the internal collapse almost vanishes
(cf.\ fig.\ \ref{fig:slope_size}). Therefore, in the following we will only consider this term to show
that we approximately obtain a parabolic behavior. Let us assume that the
$h_i$ are linear in $i$:
\begin{eqnarray}
  \label{eq:linear_hi}
  h_i=\left(1-a\frac{i}{n}\right) h_0, \qquad a<1
\end{eqnarray}
This means that deeper in the system (larger $i$) the width of the slice is
smaller. Note that for the case $a\ll 1$ this can be understood as a linearization. This
case means that the overall compaction is not large
as it is the case for intermediate Bond numbers.  
From eqs.\ (\ref{eq:linear_hi}) and (\ref{slice_simple}) we obtain:
\begin{eqnarray}
  \label{eq:parabola}
  \Delta y^{(i)}_{\rm shift}&=& \sum_{j=1}^{n-i} \left(1-a\frac{j}{n}\right)
  h_0 - \left(1-a\frac{j+i}{n}\right) h_0 \nonumber\\
&=& \sum_{j=1}^{n-i} a\frac{i}{n}h_0 = (n-i)a\frac{i}{n}h_0
\end{eqnarray}
This is a quadratic dependence on the slice number $i$. To compare to
our results we have to transform $i$ to vertical position $y_e$, which is
obtained when summing up the height $h_i$ of all slices:
\begin{eqnarray}
  \label{eq:ye_i}
 y_e(i)=\sum_{j=n}^{i+1} h_j = \sum_{j=1}^{n-i} h_{j+i}
 \end{eqnarray}
Using the approximation (\ref{eq:linear_hi}) we obtain:
\begin{eqnarray}
  \label{eq:ye_i_approx} 
 \hspace{-3ex}y_e(i) & \simeq& \sum_{j=1}^{n-i}  \left( 1-a \frac{j+i}{n} \right) h_0\\
  &=& h_0\left[ n-\frac{a}{2}(n^2-n)\right] - i h_0 \left[ 1-a(n-1/2)\right] \label{eq:ye_i_solved}
\end{eqnarray}
The detailed derivation is given in appendix \ref{sec:derive_ye_i}. 
From this equation we can obtain $i(y_e)$:
\begin{eqnarray}
  \label{eq:i_ye}
 i= \frac{-y_e}{h_0\left[ 1-a(n-1/2) \right]}+\frac{n-a/2(n^2-n)}{1-a(n-1/2)}
\end{eqnarray}
We assume that we are in the limit of relatively small $a$.  
Neglecting all terms in $a$ in eq.\ (\ref{eq:i_ye})  corresponds to
neglecting 
terms in $a^2$ in eq.\ (\ref{eq:parabola}). With this simplification
additionally using $h_0=H/n$ we obtain $i=n-y_e/h_0=n(1-y_e/H)$, leading to:
\begin{eqnarray}
  \label{eq:parabola_ye}
  \Delta y_{\rm shift}(y_e(i)) &=&  [n-i(y_e)]i(y_e)\frac{Ha}{n^2}\\
 &=& a y_e\left( 1-\frac{y_e}{H} \right) \label{eq:parabola_ye_solved}
\end{eqnarray}
This behavior is plotted in fig.\ \ref{fig:dy_Bog1e3} (green curve). Note that
in this figure the $\Delta y$ is negatively defined as opposed to the
definition used in this section.  
 This curve fits relatively well the measured curves
except coming close to the top. This can be explained by the existence of a
small ``crust'', i.e.\ an accumulation of particles at the top of the system in the simulations which is not considered in the analysis in this
section. Probably this is also the reason for the slightly different
pre-factors of the parabola: From eq.\ (\ref{eq:parabola_ye_solved}) we obtain 
$a=0.39$ leading to a pre-factor of the quadratic term of $a/H=0.00049$ which
is somewhat larger than the value obtained previously of $0.00038$. The
value of $a=0.39$ is at least reasonably  small to ensure that the
considerations of this section agree roughly with the simulation
results. Previously by scaling $\Delta y$ for different system sizes we
obtained that the pre-factor of the parabola scales as $1/H$
(cf.\ eq.\ \ref{eq:scaling}). This implies that $a$ is independent on system
size $H$, for each specific Bond number, additionally
indicating by its value how good the approximations of this section are.


 Thus, in the limit of small $a$ we
could show that the linear behavior of $\Delta y$ when collapsing after
deposition complete leads to a parabolic behavior when collapsing during
deposition. Note that this $a$ represents the differences in heights of the
top and the bottom slice, i.e.\ the assumption of small $a$ is true when the
density difference between the density close to the bottom and at the top is
small which is the case in all our cases studied here
(cf.\ Fig.\ \ref{fig:density_profiles}). In the structures studied within our
model in the previous sections (see e.g.\ sec.\ \ref{sec:depo_grav}) a small
``crust'' (particle accumulation) at the top leads to a density increase again. This will lead to a shift
in the parabolic profile to the right (to the top). As we discussed previously
the deposition/collapse process is not continuous, so that the parabola is
only an average of a very noisy distribution of $\Delta y$. Additionally, the
deposition density is not constant, but slightly increasing
(cf.\ figs.\ \ref{fig:calculation_byslides1e3} and
\ref{fig:calculation_byslides1e2}) accompanied by relatively large
fluctuations. For these reasons we can only expect a rough matching of our
theory with the simulations. Nevertheless the parabolic behavior has been
observed relatively clearly.

\section{Conclusion/Outlook}

We studied the generation of fragile granular structures by a
deposition/collapse process. In one extreme case where the deposition is
sufficiently slow to allow the system to collapse and relax due to gravity
after the deposition of each single grain we studied the influence of the
granular Bond number on 
the density profile. For intermediate
Bond numbers the density decreases with height due to the compaction of the
powder's own weight. We studied the generation process dynamics which is
discontinuous in small avalanches. These avalanches showed a parabolic
behavior and can be used to calculate the
final density profile from the deposition density. In the other extreme case
of collapse after deposition complete we found that the density is constant
with vertical position, and that the avalanche size depends linearly on
vertical position. We could relate the parabolic behavior to the linear one by
imagining  a slice by slice deposition/collapse process. 
\rev{Note that the linear behavior investigated here for the case of ballistic
  deposition followed by a gravitational collapse will be found for all
  collapse/compaction processes of homogeneous initial structures to
  homogeneous final structures. Therefore the concept of avalanches introduced
  in this paper is of general applicability to granular structures
  collapsing due to gravity or similar forces.}

 Our results maybe directly verified experimentally, as already mentioned in the
 introduction,  e.g.\ by using a Hele Shaw
 cell  \cite{voeltz2000,voeltz2001,Vinningland07}
which can be tilted to effectively change gravity.   
To apply the model presented here more specifically,  e.g.\ for snow
compaction, more realistic microscopic properties including aging processes
would have to be used. For cake formation
processes, instead of gravity a porosity dependent drag force could be
applied. In this context an explicit consideration of the pore fluid/gas could
be 
needed. The influence of the pore fluid/gas should be in particular studied
for the 
fast compaction process presented in this paper.  



\begin{acknowledgments}
 We thank Prof.\ Dietrich Wolf for fruitful discussions and the DFG (project
 HE 2732/11-1)  for financial support.
\end{acknowledgments}
 
\appendix
\section{Relation between slopes} \label{sec:relation_linear}
The linear dependence of avalanches is found as well in $y_e$ as in $y_i$ (see
fig.\ \ref{fig:immediate_dy}). In this section the relation between the two
lines is derived in detail.
Assuming  
\begin{equation}
\Delta y (y_e)=a-by_e \mbox{ and }\Delta y
(y_i)=a'-b'y_i \label{eq:linear}
\end{equation}
the values $a'$ and $b'$ can be calculated from $a$ and $b$ as
shown in the following.
The relation between $y_i$ and $y_e$ can be written as:
\begin{eqnarray}
  \label{eq:yi_vs_ye}
y_i=y_e-\Delta y(y_e)&=& y_e-(a-by_e)\\ 
 &=& y_e (1+b) -a \\
\Longrightarrow  y_e&=& \frac{a}{1+b}+\frac{\overbrace{1}^{1+b-b}}{1+b} y_i\\ 
&=& y_i\left(1-\frac{b}{1+b}\right)+\frac{a}{1+b}\label{eq:rel_ab}
\end{eqnarray}
According to (\ref{eq:ye_relation}) and (\ref{eq:linear}) $y_e$ can be written
as: 
\begin{equation}
  \label{eq:ye_linear}
y_e=y_i+(a'-b'y_i)
\end{equation}
Comparing (\ref{eq:rel_ab}) and (\ref{eq:ye_linear}) results in:
\begin{equation}
  \label{eq:relation_linear_appendix}
b'=\frac{b}{1+b}, \quad a'=\frac{a}{1+b}
\end{equation}
From this or by a similar derivation the inverse relations can also be
obtained:
\begin{equation}
  \label{eq:relation_linear_inverse}
b=\frac{b'}{1-b'}, \quad a=\frac{a'}{1-b'}
\end{equation}

\section{Derivation of $y_e(i)$ in linear approximation for $h_i$}\label{sec:derive_ye_i}
 
Here we show the details of the derivation to obtain
eq.\ (\ref{eq:ye_i_solved}) from eq.\ (\ref{eq:ye_i_approx}):
\begin{eqnarray}
 \hspace{-3ex}
&& \hspace{-3ex} y_e(i)  \simeq \sum_{j=1}^{n-i}  \left( 1-a \frac{j+i}{n}
 \right) h_0 \nonumber \\
 &=& h_0(n-i) -\frac{ha}{n}\left( \underbrace{\sum_{j=1}^{n-i}i}_{(n-i)i} +  \underbrace{\sum_{j=1}^{n-i}j}_{(n-i+1)(n-i)/2}  \right) \nonumber\\ 
 &=& h_0 (n-i) -h_0 a (n-i) i/n - h_0a (n-i+1) (n-i)/2 \nonumber\\ 
 &=&  h_0 (n-i) -h_0 a \left[ \frac{n}{2}(n+1) - i \left( (n+1)/2+n/2 \right) \right]  \nonumber\\ 
  &=& h_0\left[ n-\frac{a}{2}(n^2-n)\right] - i h_0 \left[ 1-a(n-1/2)\right] \nonumber 
\end{eqnarray}

\vfill

\bibliography{doc_density_profiles_refs}



\end{document}